\def\temp{1.34}%
\let\tempp=\relax
\expandafter\ifx\csname psboxversion\endcsname\relax
  \message{PSBOX(\temp) loading}%
\else
    \ifdim\temp cm>\psboxversion cm
      \message{PSBOX(\temp) loading}%
    \else
      \message{PSBOX(\psboxversion) is already loaded: I won't load
        PSBOX(\temp)!}%
      \let\temp=\psboxversion
      \let\tempp= 
    \fi
\fi
\tempp
\let\psboxversion=\temp
\catcode`\@=11
%
%
\def\psfortextures{
\def\PSspeci@l##1##2{%
\special{illustration ##1\space scaled ##2}%
}}%
\def\psfordvitops{
\def\PSspeci@l##1##2{%
\special{dvitops: import ##1\space \the\drawingwd \the\drawinght}%
}}%
\def\psfordvips{
\def\PSspeci@l##1##2{%
\d@my=0.1bp \d@mx=\drawingwd \divide\d@mx by\d@my
\includegraphics{##1\space}}}%
\def\psforoztex{
\def\PSspeci@l##1##2{%
\special{##1 \space
      ##2 1000 div dup scale
      \number-\psllx\space \number-\pslly\space translate
}}}%
\def\psfordvitps{
\def\psdimt@n@sp##1{\d@mx=##1\relax\edef\psn@sp{\number\d@mx}}
\def\PSspeci@l##1##2{%
\special{dvitps: Include0 "psfig.psr"}
\psdimt@n@sp{\drawingwd}
\special{dvitps: Literal "\psn@sp\space"}
\psdimt@n@sp{\drawinght}
\special{dvitps: Literal "\psn@sp\space"}
\psdimt@n@sp{\psllx bp}
\special{dvitps: Literal "\psn@sp\space"}
\psdimt@n@sp{\pslly bp}
\special{dvitps: Literal "\psn@sp\space"}
\psdimt@n@sp{\psurx bp}
\special{dvitps: Literal "\psn@sp\space"}
\psdimt@n@sp{\psury bp}
\special{dvitps: Literal "\psn@sp\space startTexFig\space"}
\special{dvitps: Include1 "##1"}
\special{dvitps: Literal "endTexFig\space"}
}}%
\def\psfordvialw{
\def\PSspeci@l##1##2{
\special{language "PostScript",
position = "bottom left",
literal "  \psllx\space \pslly\space translate
  ##2 1000 div dup scale
  -\psllx\space -\pslly\space translate",
include "##1"}
}}%
\def\psforptips{
\def\PSspeci@l##1##2{{
\d@mx=\psurx bp
\advance \d@mx by -\psllx bp
\divide \d@mx by 1000\multiply\d@mx by \xscale
\incm{\d@mx}
\let\tmpx\dimincm
\d@my=\psury bp
\advance \d@my by -\pslly bp
\divide \d@my by 1000\multiply\d@my by \xscale
\incm{\d@my}
\let\tmpy\dimincm
\d@mx=-\psllx bp
\divide \d@mx by 1000\multiply\d@mx by \xscale
\d@my=-\pslly bp
\divide \d@my by 1000\multiply\d@my by \xscale
\at(\d@mx;\d@my){\special{ps:##1 x=\tmpx, y=\tmpy}}
}}}%
\def\psonlyboxes{
\def\PSspeci@l##1##2{%
\at(0cm;0cm){\boxit{\vbox to\drawinght
  {\vss\hbox to\drawingwd{\at(0cm;0cm){\hbox{({\tt##1})}}\hss}}}}
}}%
\def\psloc@lerr#1{%
\let\savedPSspeci@l=\PSspeci@l%
\def\PSspeci@l##1##2{%
\at(0cm;0cm){\boxit{\vbox to\drawinght
  {\vss\hbox to\drawingwd{\at(0cm;0cm){\hbox{({\tt##1}) #1}}\hss}}}}
\let\PSspeci@l=\savedPSspeci@l
}}%
%
%
\newread\pst@mpin
\newdimen\drawinght\newdimen\drawingwd
\newdimen\psxoffset\newdimen\psyoffset
\newbox\drawingBox
\newcount\xscale \newcount\yscale \newdimen\pscm\pscm=1cm
\newdimen\d@mx \newdimen\d@my
\newdimen\pswdincr \newdimen\pshtincr
\let\ps@nnotation=\relax
{\catcode`\|=0 |catcode`|\=12 |catcode`|
|catcode`#=12 |catcode`*=14
|xdef|backslashother{\}*
|xdef|percentother{
|xdef|tildeother{~}*
|xdef|sharpother{#}*
}%
\def\R@moveMeaningHeader#1:->{}%
\def\uncatcode#1{%
\edef#1{\expandafter\R@moveMeaningHeader\meaning#1}}%
\def\execute#1{#1}
\def\psm@keother#1{\catcode`#112\relax}
\def\executeinspecs#1{%
\execute{\begingroup\let\do\psm@keother\dospecials\catcode`\^^M=9#1\endgroup}}%
\def\@mpty{}%
\def\matchexpin#1#2{
  \fi%
  \edef\tmpb{{#2}}%
  \expandafter\makem@tchtmp\tmpb%
  \edef\tmpa{#1}\edef\tmpb{#2}%
  \expandafter\expandafter\expandafter\m@tchtmp\expandafter\tmpa\tmpb\endm@tch%
  \if\match%
}%
\def\matchin#1#2{%
  \fi%
  \makem@tchtmp{#2}%
  \m@tchtmp#1#2\endm@tch%
  \if\match%
}%
\def\makem@tchtmp#1{\def\m@tchtmp##1#1##2\endm@tch{%
  \def\tmpa{##1}\def\tmpb{##2}\let\m@tchtmp=\relax%
  \ifx\tmpb\@mpty\def\match{YN}%
  \else\def\match{YY}\fi%
}}%
\def\incm#1{{\psxoffset=1cm\d@my=#1
 \d@mx=\d@my
  \divide\d@mx by \psxoffset
  \xdef\dimincm{\number\d@mx.}
  \advance\d@my by -\number\d@mx cm
  \multiply\d@my by 100
 \d@mx=\d@my
  \divide\d@mx by \psxoffset
  \edef\dimincm{\dimincm\number\d@mx}
  \advance\d@my by -\number\d@mx cm
  \multiply\d@my by 100
 \d@mx=\d@my
  \divide\d@mx by \psxoffset
  \xdef\dimincm{\dimincm\number\d@mx}
}}%
%
\newif\ifNotB@undingBox
\newhelp\PShelp{Proceed: you'll have a 5cm square blank box instead of
your graphics (Jean Orloff).}%
\def\s@tsize#1 #2 #3 #4\@ndsize{
  \def\psllx{#1}\def\pslly{#2}%
  \def\psurx{#3}\def\psury{#4}
  \ifx\psurx\@mpty\NotB@undingBoxtrue
  \else
    \drawinght=#4bp\advance\drawinght by-#2bp
    \drawingwd=#3bp\advance\drawingwd by-#1bp
  \fi
  }%
\def\sc@nBBline#1:#2\@ndBBline{\edef\p@rameter{#1}\edef\v@lue{#2}}%
\def\g@bblefirstblank#1#2:{\ifx#1 \else#1\fi#2}%
{\catcode`\%=12
\xdef\B@undingBox{
\def\ReadPSize#1{
 \readfilename#1\relax
 \let\PSfilename=\lastreadfilename
 \openin\pst@mpin=#1\relax
 \ifeof\pst@mpin \errhelp=\PShelp
   \errmessage{I haven't found your postscript file (\PSfilename)}%
   \psloc@lerr{was not found}%
   \s@tsize 0 0 142 142\@ndsize
   \closein\pst@mpin
 \else
   \if\matchexpin{\GlobalInputList}{, \lastreadfilename}%
   \else\xdef\GlobalInputList{\GlobalInputList, \lastreadfilename}%
     \immediate\write\psbj@inaux{\lastreadfilename,}%
   \fi%
   \loop
     \executeinspecs{\catcode`\ =10\global\read\pst@mpin to\n@xtline}%
     \ifeof\pst@mpin
       \errhelp=\PShelp
       \errmessage{(\PSfilename) is not an Encapsulated PostScript File:
           I could not find any \B@undingBox: line.}%
       \edef\v@lue{0 0 142 142:}%
       \psloc@lerr{is not an EPSFile}%
       \NotB@undingBoxfalse
     \else
       \expandafter\sc@nBBline\n@xtline:\@ndBBline
       \ifx\p@rameter\B@undingBox\NotB@undingBoxfalse
         \edef\t@mp{%
           \expandafter\g@bblefirstblank\v@lue\space\space\space}%
         \expandafter\s@tsize\t@mp\@ndsize
       \else\NotB@undingBoxtrue
       \fi
     \fi
   \ifNotB@undingBox\repeat
   \closein\pst@mpin
 \fi
\message{#1}%
}%
%
%
\def\psboxto(#1;#2)#3{\vbox{%
   \ReadPSize{#3}%
   \advance\pswdincr by \drawingwd
   \advance\pshtincr by \drawinght
   \divide\pswdincr by 1000
   \divide\pshtincr by 1000
   \d@mx=#1
   \ifdim\d@mx=0pt\xscale=1000
         \else \xscale=\d@mx \divide \xscale by \pswdincr\fi
   \d@my=#2
   \ifdim\d@my=0pt\yscale=1000
         \else \yscale=\d@my \divide \yscale by \pshtincr\fi
   \ifnum\yscale=1000
         \else\ifnum\xscale=1000\xscale=\yscale
                    \else\ifnum\yscale<\xscale\xscale=\yscale\fi
              \fi
   \fi
   \divide\drawingwd by1000 \multiply\drawingwd by\xscale
   \divide\drawinght by1000 \multiply\drawinght by\xscale
   \divide\psxoffset by1000 \multiply\psxoffset by\xscale
   \divide\psyoffset by1000 \multiply\psyoffset by\xscale
   \global\divide\pscm by 1000
   \global\multiply\pscm by\xscale
   \multiply\pswdincr by\xscale \multiply\pshtincr by\xscale
   \ifdim\d@mx=0pt\d@mx=\pswdincr\fi
   \ifdim\d@my=0pt\d@my=\pshtincr\fi
   \message{scaled \the\xscale}%
 \hbox to\d@mx{\hss\vbox to\d@my{\vss
   \global\setbox\drawingBox=\hbox to 0pt{\kern\psxoffset\vbox to 0pt{%
      \kern-\psyoffset
      \PSspeci@l{\PSfilename}{\the\xscale}%
      \vss}\hss\ps@nnotation}%
   \global\wd\drawingBox=\the\pswdincr
   \global\ht\drawingBox=\the\pshtincr
   \global\drawingwd=\pswdincr
   \global\drawinght=\pshtincr
   \baselineskip=0pt
   \copy\drawingBox
 \vss}\hss}%
  \global\psxoffset=0pt
  \global\psyoffset=0pt
  \global\pswdincr=0pt
  \global\pshtincr=0pt 
  \global\pscm=1cm 
}}%
%
%
\def\psboxscaled#1#2{\vbox{%
  \ReadPSize{#2}%
  \xscale=#1
  \message{scaled \the\xscale}%
  \divide\pswdincr by 1000 \multiply\pswdincr by \xscale
  \divide\pshtincr by 1000 \multiply\pshtincr by \xscale
  \divide\psxoffset by1000 \multiply\psxoffset by\xscale
  \divide\psyoffset by1000 \multiply\psyoffset by\xscale
  \divide\drawingwd by1000 \multiply\drawingwd by\xscale
  \divide\drawinght by1000 \multiply\drawinght by\xscale
  \global\divide\pscm by 1000
  \global\multiply\pscm by\xscale
  \global\setbox\drawingBox=\hbox to 0pt{\kern\psxoffset\vbox to 0pt{%
     \kern-\psyoffset
     \PSspeci@l{\PSfilename}{\the\xscale}%
     \vss}\hss\ps@nnotation}%
  \advance\pswdincr by \drawingwd
  \advance\pshtincr by \drawinght
  \global\wd\drawingBox=\the\pswdincr
  \global\ht\drawingBox=\the\pshtincr
  \global\drawingwd=\pswdincr
  \global\drawinght=\pshtincr
  \baselineskip=0pt
  \copy\drawingBox
  \global\psxoffset=0pt
  \global\psyoffset=0pt
  \global\pswdincr=0pt
  \global\pshtincr=0pt 
  \global\pscm=1cm
}}%
%
\def\psbox#1{\psboxscaled{1000}{#1}}%
\newif\ifn@teof\n@teoftrue
\newif\ifc@ntrolline
\newif\ifmatch
\newread\j@insplitin
\newwrite\j@insplitout
\newwrite\psbj@inaux
\immediate\openout\psbj@inaux=psbjoin.aux
\immediate\write\psbj@inaux{\string\joinfiles}%
\immediate\write\psbj@inaux{\jobname,}%
%
%
\def\toother#1{\ifcat\relax#1\else\expandafter%
  \toother@ux\meaning#1\endtoother@ux\fi}%
\def\toother@ux#1 #2#3\endtoother@ux{\def\tmp{#3}%
  \ifx\tmp\@mpty\def\tmp{#2}\let\next=\relax%
  \else\def\next{\toother@ux#2#3\endtoother@ux}\fi%
\next}%
%
%
\let\readfilenamehook=\relax
\def\re@d{\expandafter\re@daux}
\def\re@daux{\futurelet\nextchar\stopre@dtest}%
\def\re@dnext{\xdef\lastreadfilename{\lastreadfilename\nextchar}%
  \afterassignment\re@d\let\nextchar}%
\def\stopre@d{\egroup\readfilenamehook}%
\def\stopre@dtest{%
  \ifcat\nextchar\relax\let\nextread\stopre@d
  \else
    \ifcat\nextchar\space\def\nextread{%
      \afterassignment\stopre@d\chardef\nextchar=`}%
    \else\let\nextread=\re@dnext
      \toother\nextchar
      \edef\nextchar{\tmp}%
    \fi
  \fi\nextread}%
\def\readfilename{\bgroup%
  \let\\=\backslashother \let\%=\percentother \let\~=\tildeother
  \let\#=\sharpother \xdef\lastreadfilename{}%
  \re@d}%
%
%
\xdef\GlobalInputList{\jobname}%
\def\psnewinput{%
  \def\readfilenamehook{
    \if\matchexpin{\GlobalInputList}{, \lastreadfilename}%
    \else\xdef\GlobalInputList{\GlobalInputList, \lastreadfilename}%
      \immediate\write\psbj@inaux{\lastreadfilename,}%
    \fi%
    \ps@ldinput\lastreadfilename\relax%
    \let\readfilenamehook=\relax%
  }\readfilename%
}%
\expandafter\ifx\csname @@input\endcsname\relax    
  \immediate\let\ps@ldinput=\input\def\input{\psnewinput}%
\else
  \immediate\let\ps@ldinput=\@@input
  \def\@@input{\psnewinput}%
\fi%
\def\nowarnopenout{%
 \def\warnopenout##1##2{%
   \readfilename##2\relax
   \message{\lastreadfilename}%
   \immediate\openout##1=\lastreadfilename\relax}}%
\def\warnopenout#1#2{%
 \readfilename#2\relax
 \def\t@mp{TrashMe,psbjoin.aux,psbjoint.tex,}\uncatcode\t@mp
 \if\matchexpin{\t@mp}{\lastreadfilename,}%
 \else
   \immediate\openin\pst@mpin=\lastreadfilename\relax
   \ifeof\pst@mpin
     \else
     \errhelp{If the content of this file is so precious to you, abort (ie
press x or e) and rename it before retrying.}%
     \errmessage{I'm just about to replace your file named \lastreadfilename}%
   \fi
   \immediate\closein\pst@mpin
 \fi
 \message{\lastreadfilename}%
 \immediate\openout#1=\lastreadfilename\relax}%
{\catcode`\%=12\catcode`\*=14
\gdef\splitfile#1{*
 \readfilename#1\relax
 \immediate\openin\j@insplitin=\lastreadfilename\relax
 \ifeof\j@insplitin
   \message{! I couldn't find and split \lastreadfilename!}*
 \else
   \immediate\openout\j@insplitout=TrashMe
   \message{< Splitting \lastreadfilename\space into}*
   \loop
     \ifeof\j@insplitin
       \immediate\closein\j@insplitin\n@teoffalse
     \else
       \n@teoftrue
       \executeinspecs{\global\read\j@insplitin to\spl@tinline\expandafter
         \ch@ckbeginnewfile\spl@tinline
       \ifc@ntrolline
       \else
         \toks0=\expandafter{\spl@tinline}*
         \immediate\write\j@insplitout{\the\toks0}*
       \fi
     \fi
   \ifn@teof\repeat
   \immediate\closeout\j@insplitout
 \fi\message{>}*
}*
\gdef\ch@ckbeginnewfile#1
 \def\t@mp{#1}*
 \ifx\@mpty\t@mp
   \def\t@mp{#3}*
   \ifx\@mpty\t@mp
     \global\c@ntrollinefalse
   \else
     \immediate\closeout\j@insplitout
     \warnopenout\j@insplitout{#2}*
     \global\c@ntrollinetrue
   \fi
 \else
   \global\c@ntrollinefalse
 \fi}*
\gdef\joinfiles#1\into#2{*
 \message{< Joining following files into}*
 \warnopenout\j@insplitout{#2}*
 \message{:}*
 {*
 \edef\w@##1{\immediate\write\j@insplitout{##1}}*
\w@{
\w@{
\w@{
\w@{
\w@{
\w@{
\w@{
\w@{
\w@{
\w@{
\w@{\string\input\space psbox.tex}*
\w@{\string\splitfile{\string\jobname}}*
\w@{\string\let\string\autojoin=\string\relax}*
}*
 \expandafter\tre@tfilelist#1, \endtre@t
 \immediate\closeout\j@insplitout
 \message{>}*
}*
\gdef\tre@tfilelist#1, #2\endtre@t{*
 \readfilename#1\relax
 \ifx\@mpty\lastreadfilename
 \else
   \immediate\openin\j@insplitin=\lastreadfilename\relax
   \ifeof\j@insplitin
     \errmessage{I couldn't find file \lastreadfilename}*
   \else
     \message{\lastreadfilename}*
     \immediate\write\j@insplitout{
     \executeinspecs{\global\read\j@insplitin to\oldj@ininline}*
     \loop
       \ifeof\j@insplitin\immediate\closein\j@insplitin\n@teoffalse
       \else\n@teoftrue
         \executeinspecs{\global\read\j@insplitin to\j@ininline}*
         \toks0=\expandafter{\oldj@ininline}*
         \let\oldj@ininline=\j@ininline
         \immediate\write\j@insplitout{\the\toks0}*
       \fi
     \ifn@teof
     \repeat
   \immediate\closein\j@insplitin
   \fi
   \tre@tfilelist#2, \endtre@t
 \fi}*
}%
\def\autojoin{%
 \immediate\write\psbj@inaux{\string\into{psbjoint.tex}}%
 \immediate\closeout\psbj@inaux
 \expandafter\joinfiles\GlobalInputList\into{psbjoint.tex}%
}%
%
%
%
\def\centinsert#1{\midinsert\line{\hss#1\hss}\endinsert}%
\def\psannotate#1#2{\vbox{%
  \def\ps@nnotation{#2\global\let\ps@nnotation=\relax}#1}}%
\def\pscaption#1#2{\vbox{%
   \setbox\drawingBox=#1
   \copy\drawingBox
   \vskip\baselineskip
   \vbox{\hsize=\wd\drawingBox\setbox0=\hbox{#2}%
     \ifdim\wd0>\hsize
       \noindent\unhbox0\tolerance=5000
    \else\centerline{\box0}%
    \fi
}}}%
%
\def\at(#1;#2)#3{\setbox0=\hbox{#3}\ht0=0pt\dp0=0pt
  \rlap{\kern#1\vbox to0pt{\kern-#2\box0\vss}}}%
%
\newdimen\gridht \newdimen\gridwd
\def\gridfill(#1;#2){%
  \setbox0=\hbox to 1\pscm
  {\vrule height1\pscm width.4pt\leaders\hrule\hfill}%
  \gridht=#1
  \divide\gridht by \ht0
  \multiply\gridht by \ht0
  \gridwd=#2
  \divide\gridwd by \wd0
  \multiply\gridwd by \wd0
  \advance \gridwd by \wd0
  \vbox to \gridht{\leaders\hbox to\gridwd{\leaders\box0\hfill}\vfill}}%
%
\def\fillinggrid{\at(0cm;0cm){\vbox{%
  \gridfill(\drawinght;\drawingwd)}}}%
%
%
\def\textleftof#1:{%
  \setbox1=#1
  \setbox0=\vbox\bgroup
    \advance\hsize by -\wd1 \advance\hsize by -2em}%
\def\textrightof#1:{%
  \setbox0=#1
  \setbox1=\vbox\bgroup
    \advance\hsize by -\wd0 \advance\hsize by -2em}%
\def\endtext{%
  \egroup
  \hbox to \hsize{\valign{\vfil##\vfil\cr%
\box0\cr%
\noalign{\hss}\box1\cr}}}%
%
\def\frameit#1#2#3{\hbox{\vrule width#1\vbox{%
  \hrule height#1\vskip#2\hbox{\hskip#2\vbox{#3}\hskip#2}%
        \vskip#2\hrule height#1}\vrule width#1}}%
\def\boxit#1{\frameit{0.4pt}{0pt}{#1}}%
\catcode`\@=12 
%
 \psfordvips   

\magnification=\magstep1
\parskip 6pt
\font\Bbb=msbm10
\font\Bbbs=msbm8
\textfont12=\Bbb
\scriptfont12=\Bbbs
\font\rmsmall= cmr8
\font\bfsmall= cmbx8
\font\rmone= cmr10 scaled \magstep1
\font\bfone= cmbx10 scaled \magstep1
\let\mcd=\mathchardef
\mcd\Be="7C42
\mcd\Ee="7C45
\mcd\Je="7C4A
\mcd\Pe="7C50
\mcd\Re="7C52
\mcd\Ze="7C5A
\def\un{\rmone 1}
\def\al{\alpha}
\def\ep{\epsilon}
\def\la{\lambda}
\def\bra{\langle}
\def\ket{\rangle}
\def\tih{\tilde h}
\def\bah{\bar h}
\def\S{\cal S}
\def\nea{\nearrow}
\def\sea{\searrow}
\def\e{\rm e}
\centerline{\bfone Correlations of a bound interface over a random substrate}
\bigskip
\centerline{Jo\"el De Coninck\footnote{$^{(1)}$}
{\rmsmall Centre de Recherche en Mod\'elisation Mol\'eculaire, Universit\'e de 
Mons-Hainaut, 20 Place du Parc, 7000 Mons, Belgium. 
Email: joel.de.coninck@crmm.umh.ac.be}, 
Fran\c cois Dunlop\footnote{$^{(2)}$}
{\rmsmall Laboratoire de Physique Th\'eorique et Mod\'elisation (CNRS - UMR 
8089), Universit\'e de Cergy-Pontoise, 95302 Cergy-Pontoise, France. 
Email: francois.dunlop@u-cergy.fr, thierry.huillet@u-cergy.fr}, 
Thierry Huillet$^{(2)}$}
\bigskip\noindent{\bf Abstract:} {\rmsmall The correlation function of
a one-dimensional interface over a random substrate, bound to the
substrate by a pressure term, is studied by Monte-Carlo simulation. It
is found that the height correlation $\bra h_i;h_{i+j}\ket$, averaged
over the substrate disorder, fits a form $ae^{-(j/b)^c}$ to a
surprising precision in the full range of $j$ where the correlation is
non-negligible. The exponent $c$ increases from 1.0 to 1.5 when the
interface tension is taken larger and larger.} 
\medskip\noindent {\rmsmall {\bfsmall PACS.}
      {02.50.-r }{Probability theory, stochastic processes, and statistics }   
 --- {02.50.Ng }{Distribution theory and Monte Carlo studies}
 --- {05.70.Np }{Interface and surface thermodynamics}.
} 

\bigskip\noindent{\bf 1. Introduction}\medskip\noindent
Surfaces play a very important role in many processes such as
painting, coating, oil recovery... The study of a surface is usually
based on its topography $h(x,y)$ and its chemical
constituents. Different tools exist to get this information,
each working at a different scale. The measured topography is thus a
function of the scale at which the measurements are performed. 
For instance, the topography obtained with an optical
profilometer is not the same as obtained with an Atomic Force 
Microscope.

The topography is also known to play a crucial role in wetting. 
It is well established that roughness will enhance hydrophobicity or
hydrophilicity [1, 2]. Suppose that you are
considering a hydrophilic polymer surface $h^1$, on top of which you
deposit a hydrophobic polymer, the surface of which is denoted $h$.  For a
given surface $h^1$, in practice, one is often willing to modify the
surface properties of $h$ by varying the amount $h-h^1$ of 
hydrophobic polymer. The natural question is thus to study the
link between $h^1$ and $h$. 

For simplicity, let us assume that the topography can be characterized
by a one-dimensional function $h_n$. Let $\hat h(k)$ denote its
Fourier transform, 
$$
\hat h(k)={a\over L}\sum_{n=1}^N h_n\exp(2i\pi nak)
$$
where $a$ is the sampling distance, $h_n$ is the height profile at the
$n$th position, and $Na=L$. The power spectrum density (PSD) is defined by
$PSD(k)= |\hat h(k)|^2$. Practical tools in optical scattering
[3] measure directly this spectrum. 
On the other hand, it is also possible to get information about
surfaces by studying  the autocovariance function
$$
f(j)={a\over L}\sum_{n=1}^N h_nh_{n+j}
-\Bigl({a\over L}\sum_{n=1}^N h_n\Bigr)^2
$$
Of course, whenever the measurements of the heights $h_i$ are not
precise enough, this will lead to a very bad estimate of $f(\cdot)$,
this is why it is often better to determine directly the PSD. But in
model systems or numerical simulations, it may be interesting to study
$f(\cdot)$. This is the motivation of this work.
For an SOS model over a quenched random substrate, we will study the link between the
autocovariance $f(\cdot)$ of $h$ and some characteristics of both the
substrate and the film (roughness, surface tension, volume).

Background and references to the mathematical litterature about SOS
interfaces, including cases with disordered substrates, can be found in 
review articles by Velenik [4] or Giacomin [5]. Our present work differs
from previous work with disordered substrates by the pressure term in
(1) below which controls the film thickness and leads to different issues.

\bigskip\noindent{\bf 2. Model}\medskip\noindent
A quenched one-dimensional random substrate is modelled by a sequence
of independent, exponentially distributed random variables of mean
one, the heights $h^1_i$, $i\in\Ze$. A film is then deposited and 
thermalized according to the Gibbs measure with the Abraham-Smith
[6, 7] Hamiltonian
$$
{\cal H}=J\sum_{|i-j|=1}|h_i-h_j|+K\sum h_i \eqno{(1)}
$$
at temperature $kT=1$. The film height $h_i\in\Re_+$ at point $i$ obeys
$h_i\ge h^1_i$. 
For a precise definition, one may start from a Markov chain 
with state space $\Re_+$, indexed by $i=-N/2,\dots,N/2$, with initial condition
$h_{-N/2}$ fixed to a constant independent of $N$, and transition probabilities
$$
\Pe(dh_{i+1}|h_i)=dh_{i+1}e^{-J|h_{i+1}-h_i|-Kh_{i+1}}\,\un_{h_{i+1}\ge
h^1_{i+1}}\,\Big/\,{\rm normalization}\eqno{(2)}
$$
No phase transition is expected in this one dimensional model with
short range correlations, so that equivalence of suitable ensembles will
hold, for almost every realization of the substrate $h^1$. 

In the language of statistical
mechanics, the Gibbs measure with Hamiltonian (1) over $[-N/2,N/2]$ is 
an isobaric ensemble where the film volume $\sum(h_i-h^1_i)$ is 
controlled by the pressure $K$ while being allowed to fluctuate. 
In the thermodynamic limit, local functions of the film
height will follow the same law as if obtained from a fixed volume 
ensemble where $\sum (h_i-h^1_i)$ is fixed. 

We keep the isobaric ensemble, but replace the fixed initial condition and free
final condition of the Markov chain by periodic boundary conditions.
The resulting model of a film on a quenched substrate is studied by Monte-Carlo
simulation with sub-lattice parallel
dynamics, and a heat bath algorithm. Simulations were performed for
system size up to $N=10^7$ and time up to $10^5\,$ MCS/S. A typical
window of the thermal or empirical time average
$$
\bra h_i\ket=\lim_{T\nea\infty}{1\over T}\sum_t h_i(t)\eqno{(3)}
$$
versus $i$, for one substrate sample $h^1$, is shown on Fig.~1.

\bigskip
\psboxto(13cm;0cm){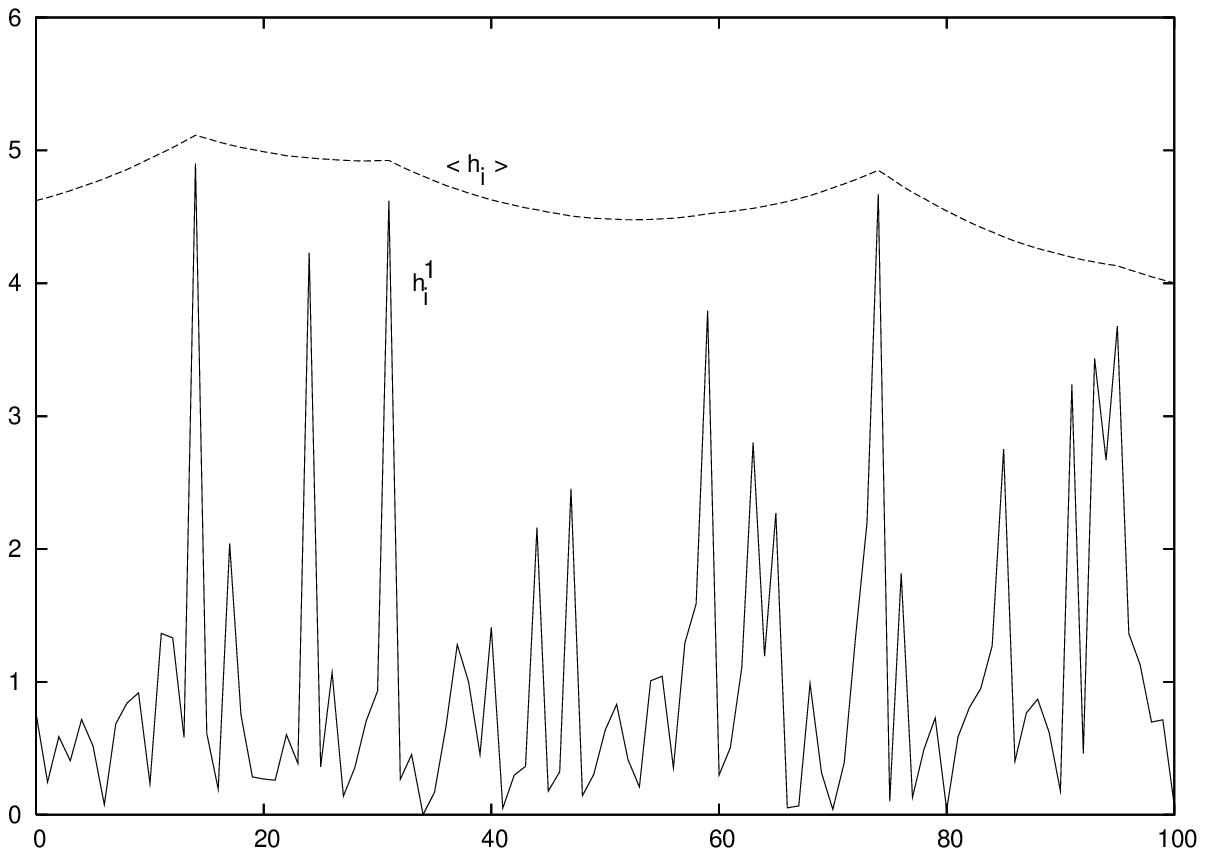}  
\vskip-0.0cm\centerline
{\rmsmall Fig.~1: Substrate $h^1_i$ and film $\bra h_i\ket$ for
$K=0.1$ and $J=10$.}
\bigskip

As may be guessed from Fig.~1, this thermal averaged profile $\bra h_i\ket$
is very close, for large $J$, to a sequence of Wulff shapes pinned at
the peaks of the substrate. This point of view was investigated by the
authors in [8]. 
In the present paper, we investigate the correlation function averaged
over the disorder, 
$$\eqalign{
f(j)&=\lim_{N\nea\infty}{1\over N}\sum_{i=-N/2}^{N/2}
\Bigl(\Ee h_ih_{i+j}-\Ee h_i\Ee h_{i+j}\Bigr)\cr
&=\lim_{N\nea\infty}{1\over N}\sum_i \bra h_i;h_{i+j}\ket\cr
&=\lim_{T\nea\infty}\lim_{N\nea\infty}{1\over T}
\sum_t\biggl[\,{1\over N}\sum_i h_ih_{i+j} 
-\Bigl({1\over N}\sum_i h_i\Bigr)^2\,\biggr]
}\eqno{(4)}
$$
where the first line is for the Markov chain (2), the second line for one
of the Gibbs measures, and the third line for our Monte-Carlo Markov
chain, which we start from $h_i=h^1_i+K^{-1}\quad\forall i$.
The second line, for the Gibbs measure in the isobaric ensemble, is
positive by the FKG inequality [9].

\bigskip\noindent{\bf 3. Numerical results}\medskip\noindent
Time averages here and below were taken after $10^4\,$MCS/S for
thermalization, during the remaining time, once in every 10 MCS/S,
altogether $T$ measurements. In the ranges of $j$ shown and discussed
below, the limits $T\nea\infty$ and $N\nea\infty$ were clearly attained.
Because of disorder effects together with thermalization time of order $L^2$
for wavelength $L$, larger $j$'s were not accessible within reasonable
computing time.

\bigskip
\psboxto(13cm;0cm){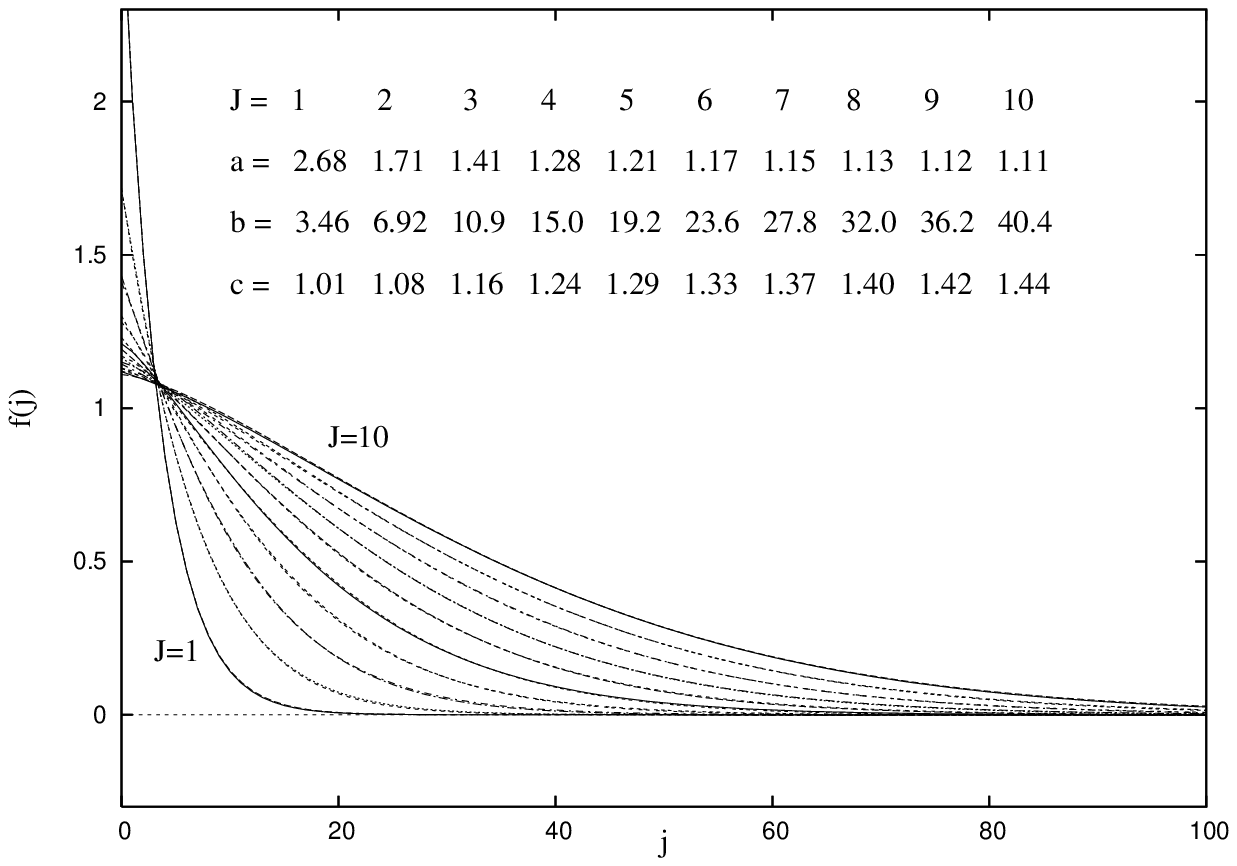}
\vskip-0.cm\centerline
{\rmsmall Fig.~2: Film correlation function $f(j)$ and fit $a\,e^{-(j/b)^c}$
for $K=0.1$ and $J=1,\dots,10$.}


At the precision of the graphs in Fig.~2, the correlation $f(j)$ and
the fitted $a\,e^{-(j/b)^c}$ are undistinguishable, over the whole range $0\le
j\le 100$, for each value of $J=1,\dots,10$. The parameter $b$ may be
considered as a correlation 
length. The exponent $c$, varying between 1 and 1.5 in our range of
parameters, shows that $f(j)$ is concave for small $j$ and convex for
large $j$, with an inflexion point at $j=b({c-1\over c})^{1/c}$.

\eject

For a given $K$, the curves for different $J$'s go through a common
point, which gives an indication of how the redundant three parameters 
$a,b,c$ of the fit depend upon the two parameters $J,K$ of the model.
Fig.~3 shows how the fixed point changes with $K$.

\bigskip
\psboxto(13cm;0cm){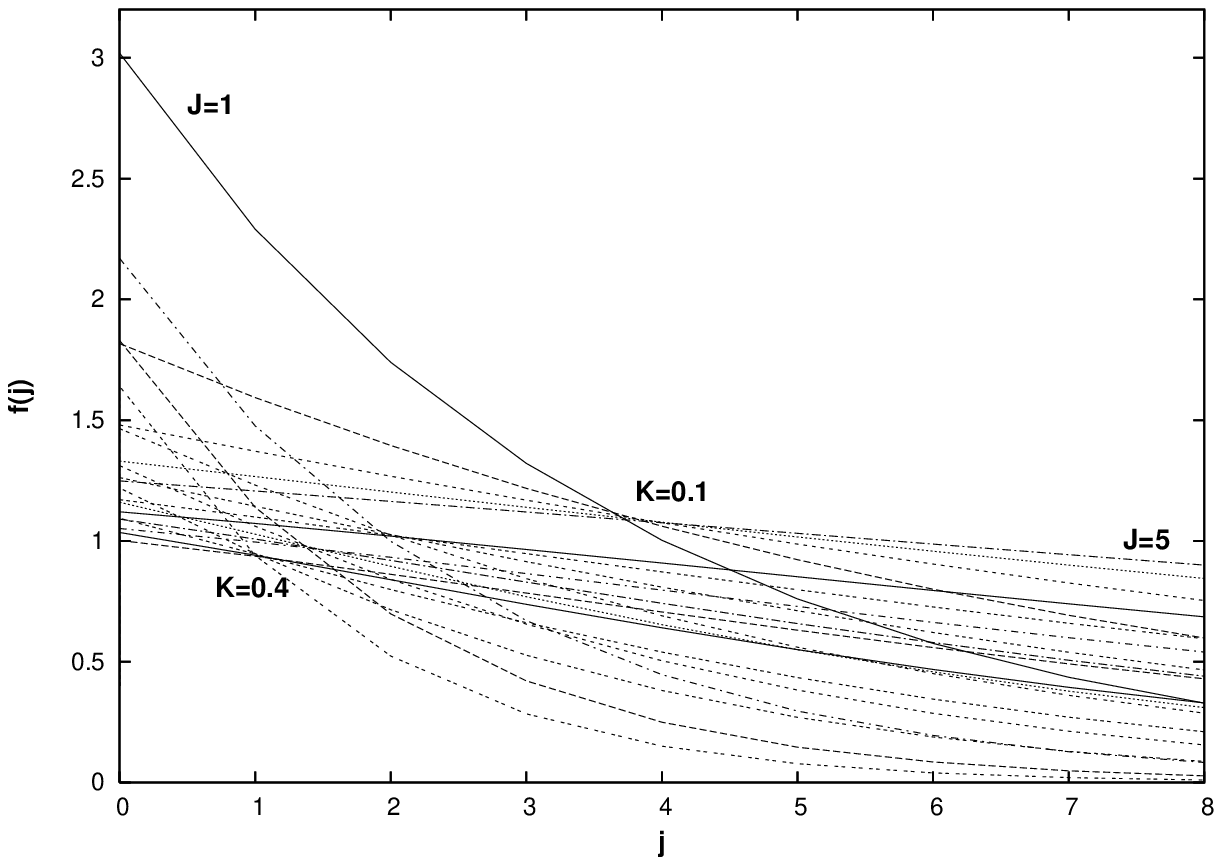}
\vskip-0cm\centerline
{\rmsmall Fig.~3: Film correlation function  $f(j)$ for
$K=0.1,0.2,0.3,0.4$ and $J=1,2,3,4,5$}
\bigskip
The super-exponential decay of the correlation function would need to
be confirmed for larger values of $j$. The fact that the disorder
makes the long-distance correlation smaller can be understood as follows. The
interval between $i$ and $i+j$ contains a family
of substrate peaks, whose number and heights go to infinity as
$j\nea\infty$. Such peaks tend to decouple the interface, because the
interface can hardly fluctuate above a peak, so that the two sides of
a peak tend to become independent. The largest peak between $i$ and
$i+j$ only makes, with large probability, a power law correction to
the flat substrate exponential decay, but the collective effect of a
density of peaks might yield super-exponential decay.

It may be, on the contrary, that exponential decay is recovered when
$j$ becomes much larger than $b$, when the correlation is too tiny to
be measured within our Monte-Carlo simulation, out of the disorder and
thermal noises.

\eject

As $J$ increases, the film tension increases, and the correlation length $b$
increases too. But the correlation function at distance $j>b$
decreases because of 
the decorrelation effect of large peaks, whose mechanism is more
effective when the surface tension is bigger: the interface tends to
be pinned at the large peaks. This results in more
``super''-exponential decay, larger exponent $c$, as appears on Fig.~4.
The top and bottom of the error bars in Fig.~4 correspond to a fit of
the correlation function in ranges $0\le j\le b$ and $j>b$ respectively.

\bigskip
\psboxto(13cm;0cm){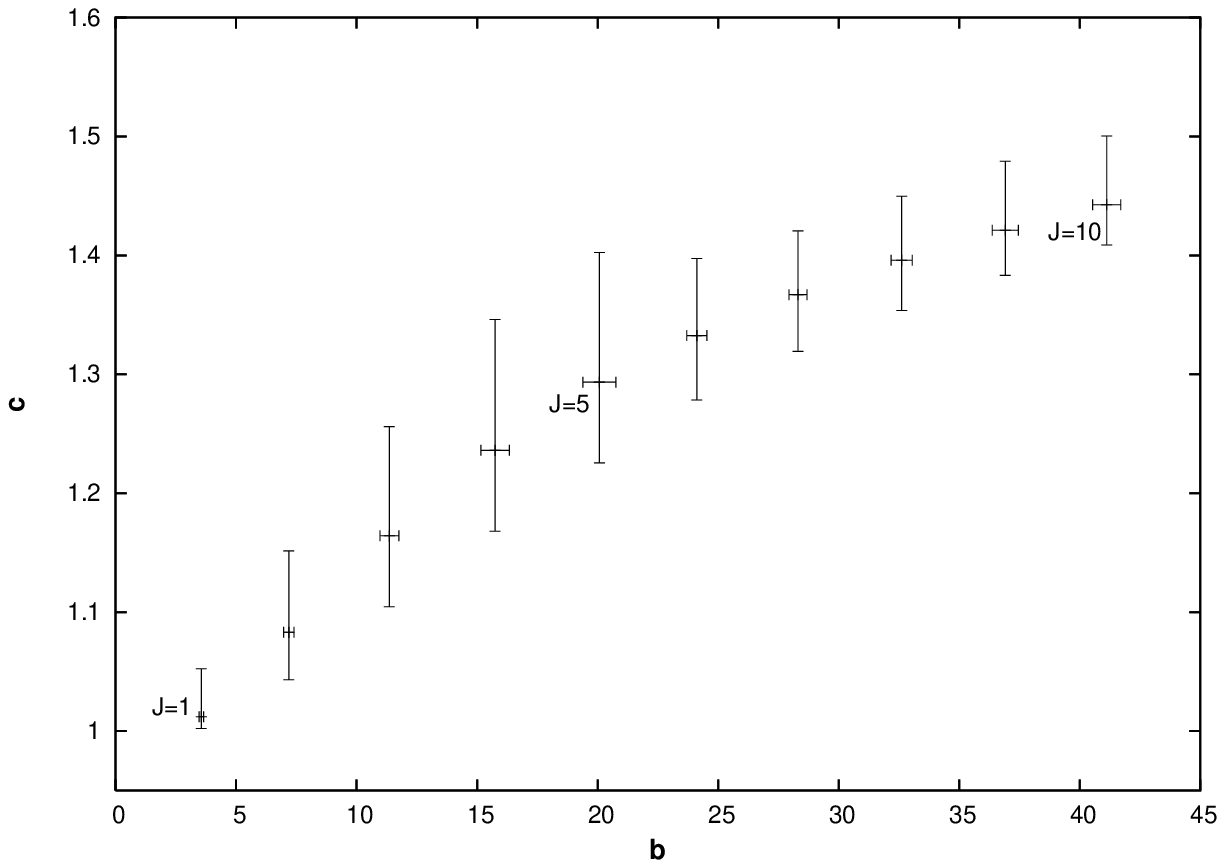}
\centerline
{\rmsmall Fig.~4: Film exponent $c$ versus correlation length $b$ for
$K=0.1$, at $J=1,\dots,10$.}

As $K$ decreases, the film volume increases. For large $K$, the film
is thin and follows closely the substrate profile, so that exponential decay
is expected. For small $K$, the film is thick, 
far from the substrate, and exponential decay is also expected.	More
precisely, one may assume that the film thickness scales when $K\sea0$
as
$K^{-1/3}$ and the correlation length $b$ as $K^{-2/3}$, like for a
flat substrate [6, 7]. For our random substrate, i.i.d. $\exp(1)$,
a peak of height $K^{-1/3}$ is likely to appear after a distance of
order $\exp(K^{-1/3})$,	which is so much larger than the correlation
length $b$ that it should affect only very weakly the exponential decay.
This is consistent with Fig. 5.

\bigskip
\psboxto(13cm;0cm){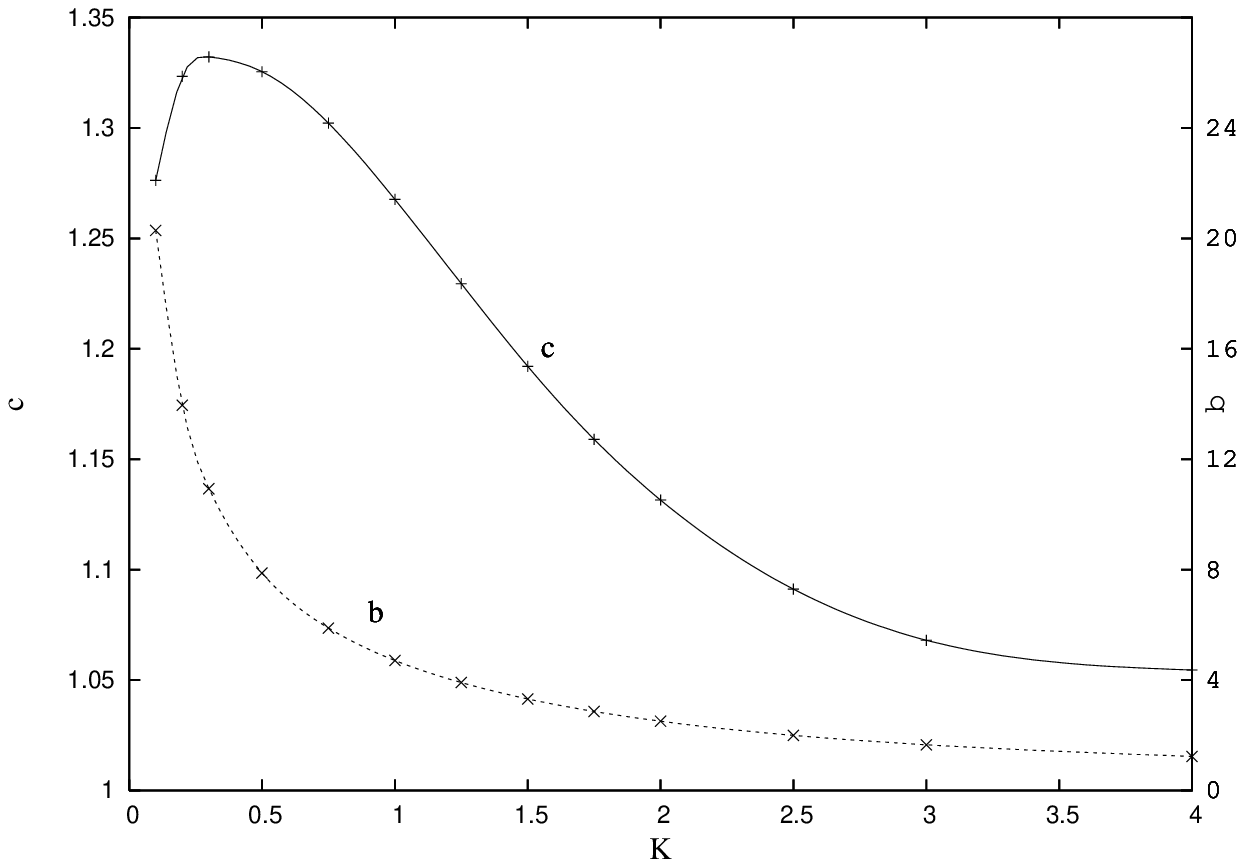}
\centerline
{\rmsmall Fig.~5: Film exponent $c$ and correlation length $b$ versus
pressure $K$ for $J=5$.}

\parskip 2pt
\medskip\noindent
{\bf References}
\smallskip\noindent

\noindent\item{1.} R. N. Wenzel: {\sl Resistance of  Solid Surfaces
to Wetting by Water}, Industrial and Engineering Chemistry, {\bf 28},
988 (1936).

\noindent\item{2.} R. N. Wenzel: {\sl Surface roughness and contact angle},
Journal of Physical and Colloid Chemistry, {\bf 53}, 1466 (1949).

\noindent\item{3.} J. C. Stover: {\sl Optical scattering : measurement and 
analysis}, SPIE Optical Engineering Press (1995).

\noindent\item{4.} Y. Velenik: {\sl Localization and Delocalization
of Random Interfaces}, Probability Surveys  {\bf 3}, 112-169 (2006).

\noindent\item{5.} G. Giacomin: {\sl Random Polymer Models},
June 2006, Imperial College Press.

\noindent\item{6.}   D.B. Abraham, E.R. Smith: {\sl Surface film
thickening: An exactly solved model}, Phys. Rev. {\bf B26}, 1480--1482, (1982).

\noindent\item{7.}  D.B. Abraham, E.R. Smith: {\sl An exactly solved 
model with a wetting transition}, J. Stat. Phys. {\bf 43}, 621--643 (1986).

\noindent\item{8.} J. De Coninck, F. Dunlop, Th. Huillet:
{\sl A necklace of Wulff shapes},  
J. Stat. Phys. {\bf 123}, 223--236 (2006).

\noindent\item{9.} C. Fortuin, P. Kasteleyn, J. Ginibre: {\sl Correlation
inequalities on some partially ordered sets}, Comm. Math. Phys. {\bf 79},
141--151 (1981).

\bye